\input{aipcheck.tex}

  \def\selectedoptions{final}

\documentclass[
   \selectedoptions
  ]
  {aipproc}

\layoutstyle{6x9}

\SetInternalRegister\hbadness{8000} 

\newcommand\doingARLO[2][]{%
  \ifx\mmref\undefined #1\else #2\fi
}
  
\begin{document}
\newcommand{\bi}{\begin{itemize}}
\newcommand{\ei}{\end{itemize}}
\newcommand{\be}{\begin{equation}}
\newcommand{\ee}{\end{equation}}
\newcommand{\ba}{\begin{eqnarray}}
\newcommand{\ea}{\end{eqnarray}}
\newcommand{\bse}{\begin{subequations}}
\newcommand{\ese}{\end{subequations}}
\newcommand{\M}{{\cal {M}}}
\newcommand{\C}{{\cal {C}}}
\newcommand{\CS}{{\cal {S}}}
\newcommand{\la}{\langle}
\newcommand{\ra}{\rangle}
\newcommand{\kB}{k_{_B}}
\newcommand{\vT}{v_{_T}}
\newcommand{\vlim}{v_{\textrm{\tiny{lim}}}}
\newcommand{\rhoha}{\rho^{\textrm{\tiny{(h)}}}}
\newcommand{\rhohac}{\rho_c^{\textrm{\tiny{(h)}}}}
\newcommand{\nha}{n^{\textrm{\tiny{(h)}}}}
\newcommand{\pha}{p^{\textrm{\tiny{(h)}}}}
\newcommand{\sha}{s^{\textrm{\tiny{(h)}}}}
\newcommand{\shac}{s_c^{\textrm{\tiny{(h)}}}}
\newcommand{\nhac}{n_c^{\textrm{\tiny{(h)}}}}
\newcommand{\xhac}{x_c^{\textrm{\tiny{(h)}}}}
\newcommand{\Tha}{T^{\textrm{\tiny{(h)}}}}
\newcommand{\bha}{\beta^{\textrm{\tiny{(h)}}}}
\newcommand{\xha}{x^{\textrm{\tiny{(h)}}}}
\newcommand{\sigha}{\sigma_{\textrm{\tiny{(h)}}}}
\newcommand{\seqq}{s^{\textrm{\tiny{(eq)}}}}
\newcommand{\neqq}{n^{\textrm{\tiny{(eq)}}}}
\newcommand{\Yeqq}{Y^{\textrm{\tiny{(eq)}}}}
\newcommand{\svir}{s_{\textrm{\tiny{(vir)}}}}
\newcommand{\zvir}{z_{\textrm{\tiny{(vir)}}}}
\newcommand{\xvir}{x_{\textrm{\tiny{(vir)}}}}
\newcommand{\xf}{x_{\textrm{f}}}
\newcommand{\nf}{n_{\textrm{f}}}
\newcommand{\sff}{s_{\textrm{f}}}
\def\chic#1{{\scriptscriptstyle #1}}
\title 
      []
      {On the thermal footsteps of Neutralino relic gases}
%
\author{Luis G. Cabral-Rosetti$^a$, Xabier Hern\'andez$^b$ and Roberto A. Sussman$^{a}$}
{address={$^a${\it Instituto de Ciencias Nucleares, Universidad Nacional 
Autónoma de México (ICN-UNAM),\\
Circuito Exterior, C.U., Apartado Postal 70-543, 04510 México, D.F.,  
México.}\\
{$^b${\it Instituto de Astronom{\'\i}a, Universidad Nacional Autónoma de México, 
(IA-UNAM).\\
Circuito de la Investigación Científica, C.U., Apartado Postal 
70-264, 04510 México, D.F.,  México.}
}\\
},
  email={luis@nuclecu.unam.mx, matias@fenix.ifisicacu.unam.mx, 
rosado@fenix.ifisicacu.unam.mx},
  thanks={}
}

\copyrightyear  {2001}

\begin{abstract}
Current literature suggests that neutralinos are the dominant cold dark 
matter particle species. Assuming the microcanonical definition of entropy, 
we examine the local entropy per particle produced between the ``freeze out'' 
era to the present. An ``entropy consistency'' criterion emerges by comparing 
this entropy with the entropy per particle of actual galactic structures 
given in terms of dynamical halo variables. We apply this criterion to the 
cases when neutralinos are mosly b-inos and mostly higgsinos, in conjunction 
with the usual ``abundance'' criterion requiring that present neutralino 
relic density complies with $0.1 < \Omega_{\chic{\tilde\chi^0_1}} < 0.3$ 
for $h\simeq 0.65$. The joint application of both criteria reveals that a 
better fitting occurs for the b-ino channels, hence the latter seem to be 
favoured over the higgsino channels. The suggested methodology can be applied 
to test other annihilation channels of the neutralino, as well as other 
particle candidates of thermal gases relics\ \footnote{Contribution to
the Proceeedings of the {\it Mexican School of Astrophysics} ({\bf EMA}), 
Guanajuato, M\'exico, July 31 - August 7, 2002 and also poster contribution
to the {\it X Mexican School of Particles and Fields}, Playa del Carmen, 
Quintana Roo, México, October 30 - November 6  2002.}
\end{abstract}


\maketitle


There are strong theoretical arguments favouring lightest supersymmetric 
particles (LSP) as making up the relic gas that forms the halos of actual 
galactic structures. Assuming that  {\it R} parity is conserved and that 
the LSP is stable, it might be an ideal candidate for cold dark matter 
(CDM), provided it is neutral and has no strong interactions. The  most 
favoured scenario \cite{Ellis,Report,Torrente,Roszkowski,Fornengo,Ellis2} 
considers the LSP to be the lightest neutralino ($\tilde\chi^0_1$), a mixture 
of supersymmetric partners of the photon, $Z$ boson and neutral Higgs boson 
\cite{Report}. Since neutralinos must have decoupled once they were 
non-relativistic, it is reasonable to assume that they constituted originaly 
a Maxwell-Boltzmann (MB) in thermal equilibrium with other components of the 
primordial cosmic plasma. In the present cosmic era, such a gas is either 
virialized in galactic halos, in the process of virialization in halos of 
galactic clusters or still in the linear regime for superclusters and 
structures near the scale of homogeneity\cite{KoTu, Padma1,Peac}. 

The equation of state of a non-relativistic MB neutralino gas is
\cite{KoTu,Padma1,Peac}
\be
\label{MBNR} \rho \ = \
m_{\chic{\tilde\chi^0_1}}\,n_{\chic{\tilde\chi^0_1}}\,
\left(1+\frac{3}{2\,x}\right),\qquad p \ = \
\frac{m_{\chic{\tilde\chi^0_1}}\,n_{\chic{\tilde\chi^0_1}}}{x}, 
\label{eqst} \qquad
 x \ \equiv \
\frac{m_{\chic{\tilde\chi^0_1}}}{T},\label{beta_def}
\ee
where $m_{\chic{\tilde\chi^0_1}}$ and $n_{\chic{\tilde\chi^0_1}}$ are the 
neutralino mass and number density.  Since we will deal exclusively with 
the lightest neutralino, we will ommit henceforth the suscript 
$_{\chic{\tilde\chi^0_1}}$, understanding that all usage of the term 
``neutralino'' and all symbols of physical and observational variables 
(\textit{i.e.} $\Omega_0,\,m,\,\rho,\,n$, etc.) will correspond to this 
specific particle. As long as the neutralino gas is in thermal equilibrium, 
we have
\ba n \ \approx \ \neqq \ =&& \
g\,\left[\frac{m}{\sqrt{2\,\pi}}\right]^3\,x^{-3/2}\,\exp\,
\left(-x\right),\label{n_theq}
\ea        
where $g=1$ is the degeneracy factor of the neutralino species. The number 
density $n$ satisfies the Boltzmann equation \cite{Report,KoTu}
\ba \dot n + 3\,H\,n \ = \ -\la
\sigma|\textrm{v}|\ra\left[n^2-\left(\neqq\right)^2\right],\label{boltz}\ea
where $H$ is the Hubble expansion factor and $\la \sigma|\textrm{v}|\ra$ is 
the annihilation cross section.  Since the neutralino is non-relativistic as
annilitation reactions ``freeze out'' and it decouples from the radiation 
dominated cosmic plasma, we can assume for $H$ and $\la\sigma|\textrm{v}|\ra$ 
the following forms
\ba H \ = \ 1.66\, g_*^{1/2}\frac{T^2}{m_p},\label{eqH}\\
\la\sigma|\textrm{v}|\ra \ = \ a \ + \ b\la \textrm{v}^2\ra,
\label{eq<sv>}\ea 
where $m_p=1.22\times 10^{19}$ GeV is Planck's mass, $g_*=g_*(T)$ is the sum 
of relativisitic degrees of freedom, $\la \textrm{v}^2\ra$ is the thermal 
averaging of the center of mass velocity (roughly $\textrm{v}^2\propto 1/x$ 
in non-relativistic conditions) and the constants $a$ and $b$ are determined 
by the parameters characterizing specific annhiliation processes of the
neutralino (s-wave or p-wave) \cite{Report}. The decoupling of the neutralino 
gas follows from the condition
\ba \Gamma \ \equiv \ n\,\la\sigma|\textrm{v}|\ra \ = \ H,\label{fcond}\ea
leading to the freeze out temperature $T_{\textrm{f}}$. Reasonable 
approximated  solutions of (\ref{fcond}) follow by solving for $x_f$ the 
implicit relation \cite{Report}
\ba \xf  = 
\ln\left[\frac{0.0764\,m_p\,c_0(2+c_0)\,(a+6\,b/\xf)\,m}{(g_{*{\textrm{f}}}
\,\xf)^{1/2}}\right],
\label{eqxf}\ea  
where $g_{*{\textrm{f}}}=g_*(T_{\textrm{f}}) $ and $c_0\approx 1/2$ yields 
the best fit to the numerical solution of (\ref{boltz}) and (\ref{fcond}). 
From the asymptotic solution of (\ref{boltz}) we obtain the present abundance 
of the relic neutralino gas \cite{Report}
\be 
\Omega_0\,h^2 \ = \ Y_\infty\, \frac{\CS_0\, m} 
{\rho_{\textrm{crit}}/h^2}
\ \approx \ 2.82\times 10^8\,Y_\infty\,\frac{m}{\textrm{GeV}},
\label{eqOmega0}
\ee

\be
Y_\infty \ \equiv \ \frac{n_0}{\CS_0}  =  
\left[0.264\,g_{*\textrm{f}}^{1/2}\,m_p\,m\left\{a/\xf+3(b-1/4\,a)
/\xf^2\right\}\right]^{-1},
\label{eqYinf}
\ee
where $\CS_0\approx 4000\,\textrm{cm}^{-3}$ is the present radiation entropy
density (CMB plus neutrinos),  $\rho_{\textrm{crit}} = 1.05 \times
10^{-5}\,\textrm{GeV}\,\textrm{cm}^{-3}$. 

Since neutralino mases are expected to be in the range of tens to hundreds of
GeV's and typicaly we have $\xf\sim 20$ so that $T_{\textrm{f}} < $ GeV, we
can use $g_{*{\textrm{f}}}\simeq 106.75$ \cite{Torrente} in equations 
(\ref{eqxf}) -- (\ref{eqYinf}). Equation (\ref{eqxf}) shows how $\xf$ has a 
logaritmic dependence on $m$, while theoretical considerations 
\cite{Ellis,Report,Torrente,Roszkowski,Fornengo,Ellis2} related to the minimal
supersymetric extensions of the Standard Model (MSSM) yield specific forms 
for $a$ and $b$ that also depend on $m$. Inserting into 
(\ref{eqOmega0})--(\ref{eqYinf}) the specific forms of $a$ and $b$ for each 
annihilation channel leads to a specific range of $m$ that satisfies the 
``abbundance'' criterion based on current observational constraints that 
require $0.1 < \Omega_0 < 0.3$ and $h\approx 0.65$ \cite{Peac}.

Suitable forms for $\la \sigma |v|\ra$ can be obtained for all types of
annihilation reactions \cite{Report}. If the neutralino is mainly pure bino, 
it will mostly annihilate into lepton pairs through t-channel exchange of 
right-handed sleptons. In this case the cross section is p-wave dominated and 
can be approximated by (\ref{eq<sv>}) with \cite{Torrente,Moroi,Olive}
\ba a \ \approx \ 0,\qquad b \ \approx
\ \frac{8\,\pi\,\alpha_1^2}{m^2\,\left[1+m_l^2/m^2\right]^2},
\label{sleptons}\ea
where $m_{l}$ is the mass of the right-handed slepton and 
$\alpha_1^2 = g_1^2/4 \pi \simeq 0.01$ is the fine structure coupling 
constant for the $U(1)_Y$ gauge interaction. If the neutralino is 
Higgsino-like, annihilating  into W-boson pairs, then the cross section is 
s-wave dominated and can be approximated by (\ref{eq<sv>}) with 
\cite{Torrente,Moroi,Olive}
\ba b \ \approx \ 0,\qquad a \ \approx \
\frac{\pi\,\alpha_2^2\,(1-m_{\chic W}^2/m^2)^{3/2}}
{2\,m^2\,(2-m_{_W}^2/m^2)^2},
\label{Wboson}\ea
where $m_{_W}=80.44$ GeV is the mass of the W-boson and 
$\alpha_2^2 = g_2^2/4 \pi \simeq 0.03$ is the fine structure coupling 
constant for the $SU(2)_L$ gauge interaction.      

In the freeze out era the entropy per particle (in units of the Boltzmann 
constant $\kB$) for the neutralino gas is given by \cite{KoTu,Peac,Padma1}
\ba \sff \ = \ \left[\frac{\rho + p}{n\,T}\right]_{\textrm{f}} \ = \
\frac{5}{2} \ + \ \xf,\label{sf}\ea
where we have assumed that chemical potential is negligible and have used
the equation of state (\ref{eqst}). From (\ref{eqxf}) and (\ref{sf}), it is 
evident that the dependence of $\sff$ on $m$ will will be determined by the 
specific details of the annihilation processes through the forms of $a$ and 
$b$. In particular, we will use (\ref{sleptons}) and (\ref{Wboson}) to compute
$\sff$ from (\ref{eqxf})-(\ref{sf}).

After decoupling, particle numbers are conserved and the neutralinos 
constitute a weakly interacting and practicaly collisionless self gravitating 
gas. This gas initialy expands with the cosmic fluid and eventualy undergoes 
gravitational clustering forming stable bound virialized structures 
\cite{Peac,Padma1,Padma2,Padma3}. The virialization process involves a 
variety of dissipative effects characterized by collisional and collisionless 
relaxation processes \cite{Padma2,Padma3,HD}.  However, instead of dealing 
with the details of this complexity, we will compare the initial and end 
states of this gas with the help of simplifying but general physical 
assumptions. 

Consider the microcanonical ensamble definition of entropy per particle for 
a diluted, non-relativistic gas of weakly interacting particles, given in 
terms of the volume of phase space \cite{Padma3}
\ba s \ = \ \ln \,\left[\frac{\,(2mE)^{3/2}\,V\,}{(2\pi\hbar)^3}\right],
\label{mcsdef}\ea
where $V$ and $E$ are local average values of volume and energy associated 
with a macrostate that is sufficiently large as to contain a large number 
of particles, but sufficiently small so that macroscopic variables are 
approximately constant. For a gas characterized by non-relativistic 
velocities $v/c\ll 1 $, we have  
$V\propto 1/n\propto m/\rho$ and $E\propto m\,v^2/2\propto m/x$. Assuming as 
the initial and final states, respectively, the decoupling 
($\sff,\,\xf,\,\nf$) and the values ($\sha,\,\xha,\,\nha$) that correspond 
to a suitable halo structure, the change in entropy per particle that follows 
from (\ref{mcsdef}) is
\ba \sha \ = \ \frac{5}{2} \ + \ \xf +
\ln\,\left[\frac{\nf}{\nha}\left(\frac{\xf}{\xha}\right)^{3/2}
\right],\label{Delta_s}\ea
where we have used (\ref{sf}) to eliminate $\sff$ in terms of $\xf$. 
Considering the halo gas as a roughly spherical, inhomogeneous and 
self-gravitating system that is the end result of the evolution and 
gravitational clustering of a density perturbation at the freeze out era 
(the initial state), the microcanonical description is an excellent 
approximation for gas particles near the symmetry center of this system 
where the density enhancement is maximum but spacial gradients are 
negligible. We will consider then current halo macroscopic variables 
(the end state) as evaluated at the center of the halo: 
$\shac,\,\xhac,\,\nhac$.      

Bearing in mind that the density perturbations at the freeze ot era were 
very small ($\delta\,\nf/\nf < 10^{-4}$, \cite{KoTu,Padma1,Peac}), the 
density $\nf$ is practicaly homogeneous and so we can estimate it from the 
conservation of particle numbers: $\nf = n_0\,(1+z_{\textrm{f}})^3$, and of 
photon entropy: $g_{*\textrm{f}}\CS_{\textrm{f}} = g_{*0}\,\CS_0\,
\,(1+z_{\textrm{f}})^3$, valid from the freeze out era to the present for 
the unperturbed homogeneous background. Eliminating $ (1+z_{\textrm{f}})^3$ 
from these conservation laws yields
\be
\label{eqnf} \nf \ = \
n_0\,\frac{g_{*\textrm{f}}}{g_{*0}}\left[\frac{T_{\textrm{f}}}
{T_0^{\textrm{\tiny{CMB}}}}\right]^3
\ \simeq \ 27.3\,n_0\,
\left[\frac{x_0^{\textrm{\tiny{CMB}}}}{\xf}\right]^3\ ,
\ee

\be
\textrm{where}\quad x_0^{\textrm{\tiny{CMB}}} \ \equiv \
\frac{m}{T_0^{\textrm{\tiny{CMB}}}}
\ = \ 4.29\,\times\,10^{12}\,\frac{m}{\textrm{GeV}}
\ee
where $g_{*0}=g_*(T_0^{\textrm{\tiny{CMB}}})\simeq 3.91$ and
$T_0^{\textrm{\tiny{CMB}}}=2.7\,\textrm{K}$. Since for present day conditions 
$n_0/\nhac=\rho_0/\rhohac$ and $\rho_0=\rho_{\textrm{crit}}\,\Omega_0\,h^2 $, 
we collect the results from (\ref{eqnf}) and write (\ref{Delta_s}) as 
\be \shac = \xf + 81.60 + \ln\left[\left(\frac{m}{\textrm{GeV}}\right)^3\,
\frac{h^2\,\Omega_0}{(\xf\,\xhac)^{3/2}}\,\frac{\textrm{GeV/cm}^3}{\rhohac}\right],
\label{shalo}
\ee 
Therefore, given $m$ and a specific form of $\la\sigma|\textrm{v}|\ra$ 
associated with $a$ and $b$, the entropy per particle of the neutralino halo 
gas depends on the initial state given by $\xf$ in (\ref{eqxf}) and 
(\ref{sf}), on observable cosmological parameters $\Omega_0,\,h$ and on 
state variables associated to the halo structure. 

If the neutralino gas in present halo structures strictly satisfies MB 
statistics, the entropy per particle, $\shac$, in terms of $\rhohac=m\,\nhac$ 
and $\xhac=m\,c^2/(\kB\,\Tha_c)$, follows from  the well known 
Sackur--Tetrode entropy formula \cite{Pathria}
\be \shac \ = 
\frac{5}{2}+\ln\left[\frac{m^4\,c^3}
{\hbar^3\,(2\pi\,\xhac)^{3/2}\,\rhohac}\right]
\nonumber\\ = 94.42 +
\ln\left[\left(\frac{m}{\textrm{GeV}}\right)^4\,\left(\frac{1}
{\xha_c}\right)^{3/2}
\,\frac{\textrm{GeV/cm}^3}{\rhohac}\right]
\label{s_halo}
\ee
Such a MB gas in equilbrium is equivalent to an isothermal halo if we
identify \cite{BT} 
\ba \frac{c^2}{\xha} \ = \ \frac{\kB\,\Tha}{m} \ 
= \ \sigha^2,\label{isot_MBa}\ea
where $\sigha^2$ is the velocity dispersion (a constant for isothermal halos). 

However, an exacly isothermal halo is not a realistic model, since its total 
mass diverges and it allows for infinite particle velocities (theoreticaly 
accessible in the velocity range of the MB distribution). More realistic halo 
models follow from ``energy truncated'' (ET) distribution functions 
\cite{Padma3,BT,Katz1,Katz2,MPV} that assume a maximal ``cut off'' velocity 
(an escape velocity). Therefore, we can provide a convenient estimate of the 
halo entropy, $\shac$, from the microcanonical entropy definition 
(\ref{mcsdef}) in terms of phase space volume, but restricting this volume to 
the actual range of velocities (\textrm{i.e.} momenta) accessible to the 
central particles, that is up to a maximal escape velocity $v_e(0)$. From 
theoretical studies of dynamical and thermodynamical stability associated 
with ET distribution functions \cite{Katz1,Katz2,cohn,MPV,HM,GZ, RST} and from 
observational data for elliptic and LSB galaxies and clusters
\cite{young,DBM,HG,FDCHA1,FDCHA2}, it is reasonable to assume
\ba v_e^2(0) \ = \ 2\,|\Phi(0)| \ \simeq \ \alpha \, \sigha^2(0),\quad 12 <
\alpha < 18,\label{alphas}\ea
where $\Phi(r)$ is the newtonian gravitational potential. We have then   
\be \shac \ \simeq \
\ln\left[\frac{m^4\,v_{e}^3}
{(2\pi\hbar)^3\,\rhohac}\right]\nonumber = 89.17
+\ln\left[\left(\frac{m}
{\textrm{GeV}}\right)^4\,\left(\frac{\alpha}{\xha_c}\right)^{3/2}
\,\frac{\textrm{GeV/cm}^3} {\rhohac}
\right]\ ,
\label{SHALO}
\ee    
where we used $\xhac=c^2/\sigha^2(0)$ as in (\ref{isot_MBa}). As expected, 
the scalings of (\ref{SHALO}) are identical to those of (\ref{s_halo}). 
Similar entropy expressions for elliptic galaxies have been examined in 
\cite{LGM}.

Comparison between $\shac$ obtained from (\ref{SHALO}) and from (\ref{shalo}) 
leads to an ``entropy consistency'' criterion. Since (\ref{SHALO}) scales 
with \, $\ln\,m^4$, while (\ref{shalo}) does so approximately with 
\, $\ln\,m^3$, we have a weak logarithmic dependence of $\shac$ on $m$. 
Therefore, the fulfilment of the  ``entropy consistency'' criterion 
identifies a specific mass range for each dark matter particle. This allows 
us to discriminate, in favour or against, suggested dark matter particle 
candidates and/or annihilation channels by verifying if the standard 
abundance criterion (\ref{eqOmega0}) is simultaneously satisfied for this 
range of masses. It is interesting to notice that  both equations, 
(\ref{shalo}) and (\ref{SHALO}), display an identical functional dependence of
$\shac$ on the present day halo parameters, $\rhohac$ and $\xha_c$. This 
implies that a given dark matter particle candidate, characterized by $m$ and 
by specific annihilation channels given by $\xf$ through (\ref{eqxf}), will 
pass or fail to pass this consistency test independently of the details one 
assumes regarding the present day dark halo structure. However, the actual 
values of $\shac$ for a given halo structure, whether obtained from 
(\ref{SHALO}) or from (\ref{shalo}), do depend on the precise values of 
$\rhohac$ and $\xha_c$.   

We will now evaluate (\ref{SHALO}) and (\ref{shalo}) for the two cases of 
neutralino channels: the b-ino and higgsino, including numerical estimates for
$\xha$ and $\rhoha$ that correspond to central regions of actual halo 
structures. Considering terminal velocties in rotation curves we have 
$v_{\textrm{term}}^2\simeq 2 \sigha^2(0)$, so that 
$\xhac \simeq 2(c/v_{\textrm{term}})^2$, while recent data from LSB
galaxies and clusters \cite{FDCHA1,FDCHA2,IS1,IS2,IS3} suggest the range of 
values $0.01\,\textrm{M}_\odot/\textrm{pc}^3
< \rhohac < 1\,\textrm{M}_\odot/\textrm{pc}^3$. Hence, we will use in 
the comparison of (\ref{shalo}) and (\ref{SHALO}) the following numerical 
values: $\rhohac=0.01\,\textrm{M}_\odot/\textrm{pc}^3 =0.416
\,\textrm{GeV/cm}^3$ and $\xhac = 2\times 10^6$, typical values for a large 
elliptical or spiral galaxy with  
$v_{\textrm{term}}\simeq 300\,\textrm{km/sec}$ \cite{IS1, IS2, IS3}.
\begin{figure}
\includegraphics[height=.45\textheight]{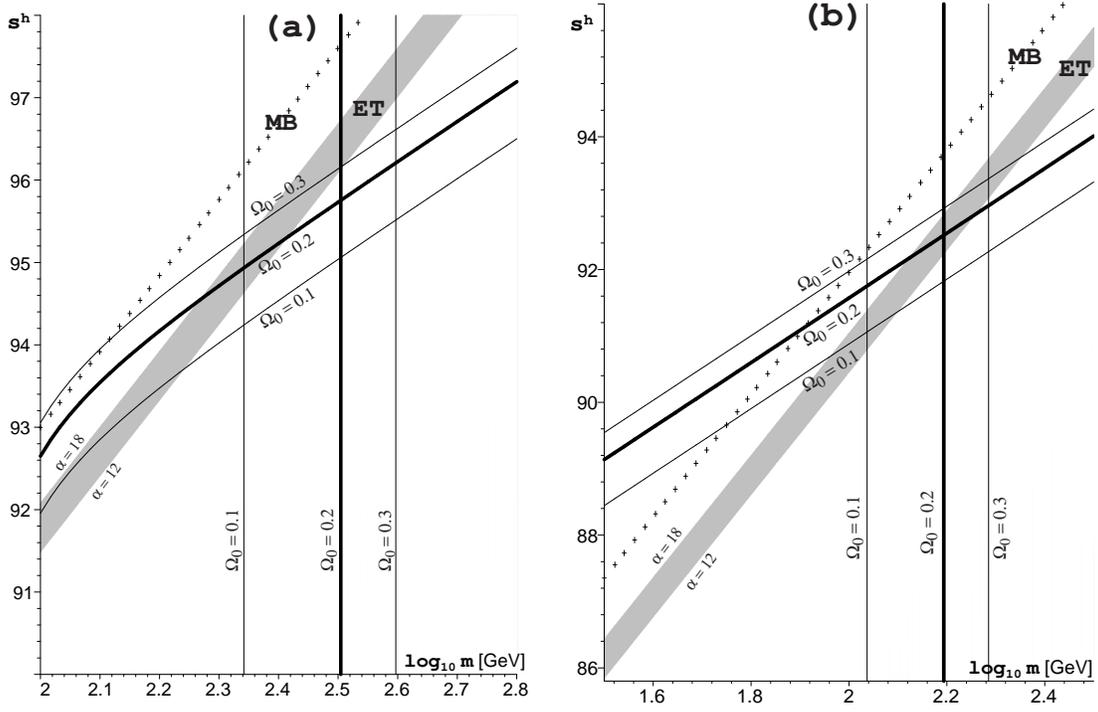}
\caption{
Figures (a) and (b) respectively correspond to the higgsino and b-ino 
channels. The figures display $\shac$ as a function of $\log_{\chic {10}} m$, 
obtained from (\ref{alphas})--(\ref{SHALO}) (gray strip), from (\ref{s_halo}) 
(crosses) and from (\ref{shalo}) for $h=0.65$ and $\Omega_0=0.2$ 
(thick curve), together with its uncertainty strip $\Omega_0=0.2\pm0.1$. The 
vertical strip marks the range of values of $m$ that follow from 
(\ref{eqOmega0})--(\ref{eqYinf}) for the same values of $\Omega_0$ and $h$. 
It is evident that only the b-ino channels allow for a simultaneous fitting 
of both the abundance and the entropy criteria.}
\label{fig1:wide}
\end{figure}
Figure 1a displays the $\shac$ as a function of $\log_{\chic {10}}\,m$, for 
the halo structure described above, for the case of a neutralino that is 
mostly higgsino. The shaded region marks $\shac$ given by (\ref{SHALO}) for 
the range of values of $\alpha$, while the vertical lines correspond to the 
range of masses selected by the abundance criterion (\ref{eqOmega0}) for 
$\Omega_0=0.1,\,0.2,\,0.3$. The solid curves are $\shac$ given by 
(\ref{shalo}) for the same values of $\Omega_0$, intersecting the shaded 
region associated with (\ref{SHALO}) at some range of masses. However, the 
ranges of coincidence of a fixed (\ref{shalo}) curve with the shaded region
(\ref{SHALO}) occurs at masses which correspond to values of $\Omega_0$ that 
are different from those used in (\ref{shalo}), that is, the vertical lines
and solid curves with same $\Omega_0$ intersect out of the shaded region. 
Hence, this annihilation channel does not seem to be favoured.

Figure 1b depicts the same variables as figure 1a, for the same halo 
structure, but for the case of a neutralino that is mostly b-ino. In  this 
case, both the abundance and the entropy criterion yield consistent mass 
ranges, which  allows us to favour this annihilation channel as a plausible 
dark matter candidate, with $m$ lying in the narow ranges given by this 
figure for any chosen value of $\Omega_0$. As noted above, the results of 
figures 1a and 1b are totaly insensitive to the values of halo variables, 
$\xhac$ and $\rhohac$, used in evaluating (\ref{SHALO}) and (\ref{shalo}). 
Different values of these variables (say, for a different halo structure) 
would only result in a relabeling of the values of $\shac $ along the 
vertical axis of the figures. 

We have presented a robust consistency criterion that can be verified for any
annihilation channel of a given dark matter candidate proposed as the
constituent particle of the present galactic dark matter halos. Since we 
require that $\shac$ of present dark matter haloes must match $\shac$ derived 
from the microcanonical definition and from freeze out conditions for the 
candidate particle, the criterion is of a very general applicability, as it 
is largely insensitive to the details of the structure formation scenario 
assumed. Further, the details of the present day halo structure enter only 
through an integral feature of the dark halos, the central escape velocity, 
thus our results are also insensitive to the fine details concerning the 
central density and the various models describing the structure of dark 
matter halos. A crucial feature of this criterion is its direct dependence 
on the physical details (\textit{i.e.} annihilation channels and mass) of 
any particle candidate.  

We have examined the specific case of the lightest neutralino for the mostly
b-ino and mostly higgsino channels. The joint application of the ``entropy 
consistency'' and the usual abbundance criteria clearly shows that the b-ino 
channel is favoured over the higgsino.  This result can be helpful in 
enhancing the study of the parameter space of annihilation channels of LSP's 
in  MSSM models, as the latter only use equations (\ref{eqxf}) and 
(\ref{eqOmega0})--(\ref{eqYinf}) in order to find out which parameters yield 
relic gas abbundances that are compatible with observational constraints
\cite{Ellis,Report,Torrente, Roszkowski,Fornengo,Ellis2}. However, equations 
(\ref{eqxf}) and (\ref{eqOmega0})--(\ref{eqYinf}) by themselves are 
insufficient to discriminate between annihilation channels. A more efficient 
study of the parameter space of MSSM can be achieved by the joint usage of 
the two criteria, for example, by considering more general cross section 
terms (see for example \cite{Report}) than the simplified approximated forms 
(\ref{sleptons}) and (\ref{Wboson}). This work is currently in progress.\\

\begin{theacknowledgments}
This work has been supported in part by {\bf DGAPA-UNAM} grant, under project 
No. {\tt IN109001} and in part by {\bf CoNaCyT} grants, under projects No. 
{\tt I37307-E} and No. {\tt I39181-E}. R.A.S. acknowledges inspiration from 
the wise meowing of his feline friends Moquis and Tontita.
\end{theacknowledgments}

\end{document}